\documentclass[preprints,article,accept,moreauthors,pdftex]{Definitions/mdpi}

\firstpage{1}
\pubvolume{1}
\issuenum{1}
\articlenumber{0}
\pubyear{2022}
\copyrightyear{2022}
\datereceived{}
\dateaccepted{}
\datepublished{}
\hreflink{https://doi.org/} % If needed use \linebreak
\Title{Experimental and Theoretical Study of Solitary-like Wave Dynamics of Liquid Film Flows over a Vibrated Inclined Plane}

\TitleCitation{Experimental and theoretical study of solitary-like wave dynamics of liquid film flows over a vibrated inclined plane}

\definecolor{my_green}{rgb}{0.55, 0.71, 0.0}

\newcommand{\Rey}{{\mathcal Re}}% Reynolds number

\usepackage{textcomp}
\usepackage{bm}
\usepackage{comment}

\Author{Ivan S.~Maksymov $^{1}$*\orcidA{} and Andrey Pototsky $^{2}$\orcidB{}}
\AuthorNames{Ivan S.~Maksymov and Andrey Pototsky}

\AuthorCitation{Maksymov, I.S.; Pototsky, A.}
\address{%
$^{1}$ \quad Optical Sciences Centre, Swinburne University of
Technology, Hawthorn, VIC 3122, Australia; imaksymov@swin.edu.au; Tel.: +61-3-3921-4805\\
$^{2}$ \quad Department of Mathematics, Swinburne University of
Technology, Hawthorn, VIC 3122, Australia; apototskyy@swin.edu.au; Tel.: +61-3-9214-4653}

\abstract{Solitary-like surface waves that originate from the spatio-temporal evolution of falling liquid films have been the subject of theoretical and experimental research due to their unique properties that are not readily observed in the physical system of other nature. Here we investigate, experimentally and theoretically, the dynamics of solitary-like surface waves in a liquid layer on an inclined plane that is subjected to a harmonic low-frequency vibration in the range from $30$ to $50$\,Hz. We demonstrate that the vibration results in a decrease in the average and peak amplitude of the long solitary-like surface waves. However, the speed of these waves remains largely unaffected by the vibration, implying that they may propagate over large distances almost without changing their amplitude, thus rendering them suitable for a number of practical applications, where the immunity of pulses that carry information to external vibrations is required.}

\keyword{falling liquid films; solitary waves; surface waves; vibrations}

\begin{document}

\section{Introduction\label{sec:1}}

Solitary waves---physical waves that maintain their shape and move with a constant velocity due to a cancellation of nonlinear effects and dispersive processes in the medium \cite{Rem94}---have been a long-term subject of fundamental and applied research studies in the fields of optics \cite{Kiv03}, fluid dynamics \cite{Kor95}, magnetism \cite{Sco05}, acoustics \cite{Per17}, electronics \cite{Xia09} and biology \cite{Hei05, Gon14}. However, despite a good understanding of the physical properties of solitary waves of different kinds, their experimental studies often involve expensive and difficult to operate equipment such as intense laser beams and nonlinear-optical materials in the field of optics \cite{Kiv03} and sources of high-power microwave radiation in the field of magnetism \cite{Sco05}, respectively. Yet, in some systems such as biological nerve fibres \cite{Hei05, Gon14} the observation of solitary-like waves requires significant preparatory works and is possible mostly when a number of specific experimental conditions are satisfied. Such technical challenges complicate both fundamental studies and verification of numerous theoretical works predicting that solitary waves could be used in communication \cite{Hau96, Cor20}, sensing \cite{Kul12} and data processing \cite{Sil21} devices and systems.

There also exists a class of material solitary-like surface waves that originate from spatio-temporal evolution of falling liquid films \cite{Cha94, Kal12}. Since the equipment needed to create falling liquid films is, in general, simpler than that used in experiments in the fields of optics and magnetism, the waves of this kind have attracted attention of many scientists \cite{Yih63, Ben66, Shk71, Por72, Nak75, Siv80, Pum83, Ale85, Tri91, Gol94, Duk95, Oro97, Ngu00, Thi04} following the pioneering experiments conducted by the Kapitzas \cite{Kap49}. In fact, while such solitary-like surface waves share many physical features with the other known types of solitary waves, they can exhibit unique physical properties not observed in other systems \cite{Ker94, Gol94, Vla01}. For instance, they can merge instead of passing through each other without significant change, with the latter being the case of two solitary waves governed by the well-known KdV equation \cite{Kor95, Zab65}. The analysis of solitary-like surface waves in flowing liquid films is also important because liquid films, as well as similar physical systems \cite{Bro70, Bal04, Liu05}, are often encountered in the fields of earth and planetary sciences \cite{Woo00, Sha21} and in technological processes \cite{Dut07}, where the liquids of interest can also experience temperature gradients \cite{Thi04, Kal12} and vibrations \cite{Gud88, Sah95, Beh14}. Given this, the effect of vibrations on the wave dynamics of film flows has become an independent subject of fundamental and applied research \cite{Woo95, Bur01, Gar13, Gar17, Zha22}. In particular, it has been shown that vibrations can suppress certain waves on the surface of flowing liquid films \cite{Woo95} but in a relevant experiment \cite{Bru07} it has been demonstrated that vibrations can promote unusual regimes of spontaneous drop movement. Speaking broadly, the study of the effect of vibrations should also help develop communication, sensing and data processing systems that are immune to undesirable mechanical impacts on devices that use liquids as a medium that provides the critical functionality (see, e.g.,~\cite{Che17, Mak17, Mak19}).

Although, traditionally, greater attention has been paid to the wave dynamics on free-falling vertical liquid films \cite{Cha94, Kal12}, studies of surface waves on liquid films flowing over slightly inclined planes have also been conducted given an essentially the same physics as in the case of vertical systems \cite{Gol94, Woo95}. However, reports on experimental results involving the effect of vibrations are rather scarce and scattered in the literature sources \cite{Gud88, Sah95, Gar17}. In particular, in \cite{Gud88} it has been shown that the vibration of a horizontal tube with a liquid thin film flowing over it results in the appearance of ripple waves at the vibration frequency. The amplitude of the so-created waves depends on the vibration amplitude and can reach the amplitude of periodic waves existing on the film surface without vibration. Subsequently, high-amplitude vibrations result in an increase in the film thickness and a concomitant increase in the speed of the waves. However, the opposite conclusions were drawn in \cite{Sah95}, which is, most likely, a result of the differences in the system (a liquid film under two-phase flow conditions) investigated in that paper.  It is also well-known that in horizontal liquid layers a harmonic vibration excites two different types of standing surface waves: harmonic waves that oscillate at the vibration frequency and subharmonic waves that oscillate at the half of the vibration frequency \cite{Kum96}. However, the presence of a mean flow across the layer changes the response frequency of the excited waves \cite{Bur01, Woo95, Gar13}. Surface waves excited by harmonic vibration in a liquid film flowing over a vertical plane were investigated experimentally in \cite{Gar17} and the results obtained in that work validated the linear theory developed in \cite{Bur01, Woo95, Gar13}.

Thus, mostly the experimental work \cite{Gar17} represents an attempt to systematically study the physics of wave motion on a vibrated plane. However, in general, building a setup involving liquids flowing down a vibrated vertical surface requires non-standard equipment built according to demanding technical specifications. In particular, the liquid should be supplied to the inlet located at the upper part of the plane so that the flow rate is not affected by the vibration. This is because the thickness of the liquid film is known to be very sensitive to external disturbances, including vibrations caused by the pump used to deliver the liquid from a reservoir to the inlet \cite{Ale85}. Moreover, the shaker producing the vibration should be connected to the vertically positioned surface via a vibration transmission structure. Some of the engineering challenges of creating such a structure are the need to move a considerable total mass of the supporting structure and liquid with high precision, and to ensure that the amplitude of the vibration across the plane area is uniform \cite{Gar17}. To resolve the problem of non-uniform vibration amplitude, in Ref.~\cite{Gar17} it is was suggested that qualitatively similar results could be obtained vibrating just one side of the plane, i.e.~vibrating just a portion of the liquid, thus also significantly reducing the total mass that needs to be moved by the shaker.

In this paper, we present and discuss a technically simple and compact experimental setup for the investigation of solitary-like surface waves on a slightly inclined plane positioned on top of a vibrating table and equipped with an auxiliary channel that recycles the liquid used in experiment, thus decreasing the chance of spills of the liquid and its unwanted contact with the measurement and imaging equipment, and also decreasing the total mass that needs to be moved by the shaker. We employ this setup to demonstrate that the instabilities of the thin liquid film caused by the vibrations result in a decrease in the peak amplitude of the solitary-like surface waves. We conclude that, despite these changes, the speed of the solitary-like waves does not appreciably change due to vibration. As a result, these waves can propagate for long distances without changing their shape and, therefore, can be used in the practical applications discussed in this work. We also demonstrate the advantage of using frequency-wavevector dispersion maps for the analysis of the properties of rolling waves, thus extending the toolbox of experimentalists working on this class of wave motion phenomena. Our experimental results are validated using the Shkadov model \cite{Shkadov67,Shkadov68}---a boundary-layer hydrodynamic model derived from the Navier-Stokes equation under the assumption of self-similar parabolic longitudinal velocity flow field across the layer.
\begin{figure}[H]
 \includegraphics[width=0.7\textwidth]{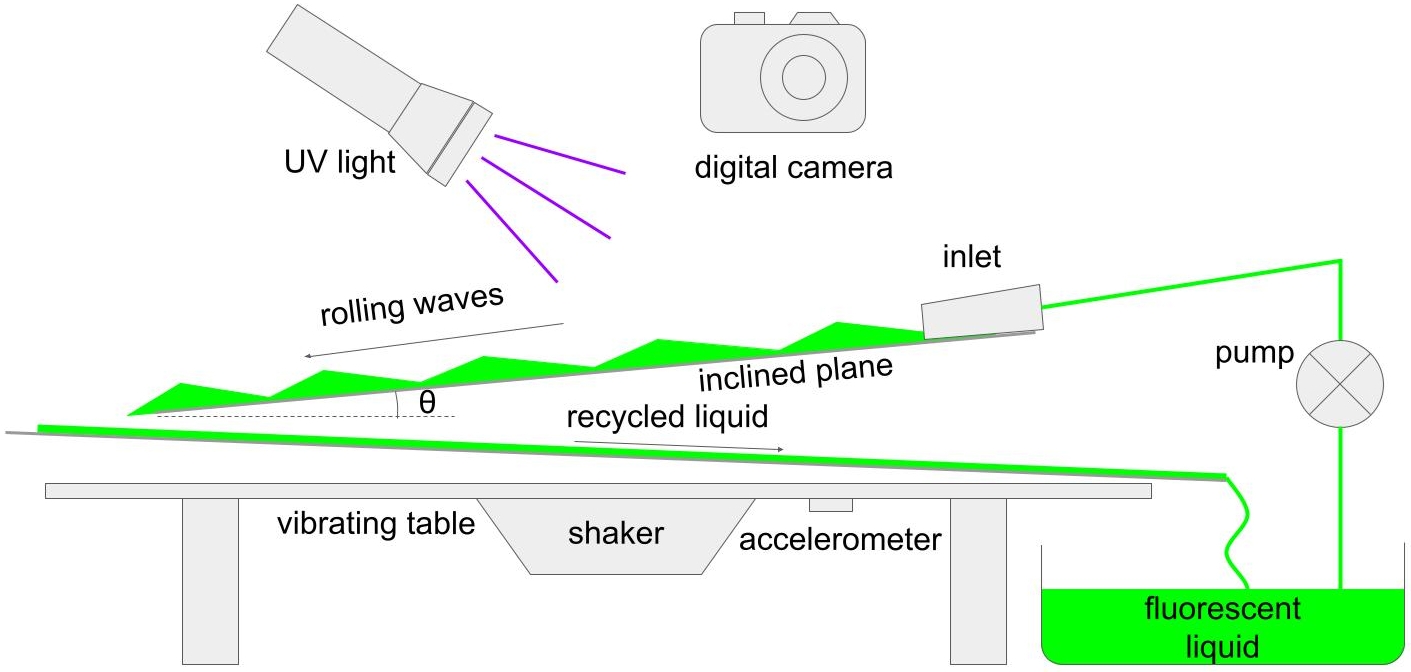}
 \caption{Sketch of the experimental setup used to observe the solitary-like surface waves. For the sake of clarity, only the main constructive features are shown, including the inclined plate, where the waves are observed, the pathway for recycling of the used liquid and the vibrating table. The dimensions and relative positions of the components in this sketch are not to scale.
 \label{Fig1}}
\end{figure}

\section{Background and Experimental Methods}
When a single-layer liquid film flows down an inclined plane with a no-slip boundary, the resulting Nusselt flat film flow profile assumes a parabolic longitudinal velocity shape having the largest velocity at the free surface \cite{Cha94, Gol94, Kal12}. In this flow regime, a long-wavelength surface instability develops when the average flow rate exceeds a certain critical value \cite{Yih63}. When the disturbances are excited naturally, in general four regimes of different wave behaviour can be observed in the downstream regions of the inclined plane \cite{Cha94}. The first regime is observed in a section of the plane that is adjacent to the inlet of the liquid, where small disturbances caused by the inlet structure are amplified while moving downstream and forming predominantly monochromatic waves. The second regime is observed in the following downstream region, where the monochromatic waves grow in amplitude and then develop higher-order frequency harmonics due to nonlinear effects. Then, as a result of complex nonlinear interactions, two-dimensional solitary-like waves are formed, and then they propagate further downstream exhibiting unique properties that, in part, coincide with those of other known solitary waves but, in general, are unique \cite{Gol94}. Finally, three-dimensional waves start to form due to transverse variations \cite{Cha94, Kal12}.

It is noteworthy that not all aforementioned regimes can necessarily be observed in practice \cite{Cha94}. Yet, it is well-known that when the initial natural disturbance at the inlet is nearly monochromatic, the waves emerging in the region located immediately after the inlet can first inherit the frequency of the disturbance and then evolve into a solitary-like wave far downstream \cite{Cha94, Gol94, Kal12}. However, when either the thickness of the liquid film or the fluid flow is periodically modulated at the inlet, solitary-like surface waves develop almost immediately after leaving the inlet area \cite{Gol94, Kal12}, which indicates that the nonlinear evolution of the flow over an inclined plane is dominated by solitary-like waves independently of whether their formation was deliberately forced or resulted naturally.

Figure~\ref{Fig1} shows a sketch of the setup that enables observing the formation of both forced and natural (unforced) solitary-like surface waves. The setup is assembled on a vibrating table that is driven by a shaker (35\,W, 20--80\,Hz response, Dayton Audio, USA) and where the vibration amplitude is controlled by an analog accelerometer (ADXL326, Analog Devices, USA). The inclined plane, where the waves are observed, is a 5-cm-wide and 50-cm-long rigid aluminium plate. The surface of the plate was chemically treated to improve the formation of the liquid film. The inclination angle is $\theta=3$\,$^{o}$. The inclined plate was mounted on top of wider open channel used to recycle the liquid by redirecting it to the main reservoir. A low-vibration DC voltage pump driven via a customised electronic circuit was used to supply water from the reservoir to the inlet. The electronic modulation of the pump flow rate enabled controlling the thickness of the liquid film and creating solitary-like waves. At the stage of preparation to the experiments, an organic fluorescent dye (Tintex, Australia) was added to tap water in the concentration of 1\,g per litre, thus leading to the emission of bright green fluorescence light when the surface of the inclined plate was illuminated with UV-A light. All experiments were conducted in a darkened room using an overhead digital camera capable of recording videos in a slow motion regime. The resulting videos were post-processed in Octave software using customised computational procedures enabling the extraction of the wave amplitude from the intensity profile of the fluorescence images. All experiments were conducted in an acoustically isolated room with environmental humidity and temperature levels.

\begin{figure}[H]
 \includegraphics[width=0.7\textwidth]{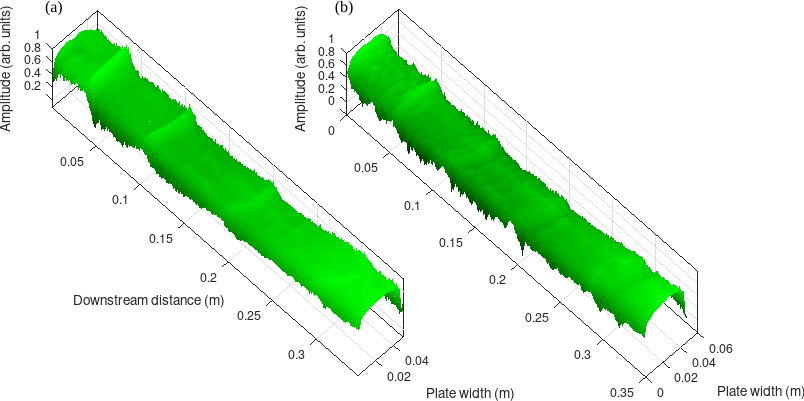}
 \caption{Instantaneous profiles the surface waves in a liquid film flowing over the inclined plate. The profiles were obtained from the fluorescence images. The frequency of the flow forcing resulting in the formation of solitary-like waves is 2\,Hz. (\textbf{a})~No vibration. (\textbf{b})~The inclined plate is vibrated with the frequency of 48\,Hz and the peak vibration amplitude of 1$g$.
 \label{Fig2}}
\end{figure}

\section{Experimental Results}
Figure~\ref{Fig2} shows the representative images of the solitatry-like surface waves propagating over a downstream section of the inclined plate, when the vibration is turned off (Fig.~\ref{Fig2}a) and when the plate is vibrated with the frequency of 48\,Hz and the peak amplitude of 1$g$ (Fig.~\ref{Fig2}b), being $g$ the gravitational acceleration. The frequency of the forcing of the solitary-like waves is 2\,Hz in both panels of Fig.~\ref{Fig2}. The images were obtained from the selected individual fluorescence frames of the recorded videos of the propagating waves. Without vibration (Fig.~\ref{Fig2}a), we can observe a train of downstream-propagating solitary pulses. A closer inspection also reveals the existence of periodic waves with an amplitude that is much smaller than that of solitary-like waves. When the plate is subjected to vibration (Fig.~\ref{Fig2}b), we continue observing a train of solitary pulses with an approximately the same pulse periodicity as in Fig.~\ref{Fig2}a. However, the peak amplitude of the pulses is lower than in the case without vibration. Yet, in agreement with the relevant theory \cite{Woo95, Gar13} and experiment on the vertical plane \cite{Gar17}, we also observe the short wavelength ripples arising due to the onset of the Faraday instability.

\begin{figure}[H]
 \includegraphics[width=0.7\textwidth]{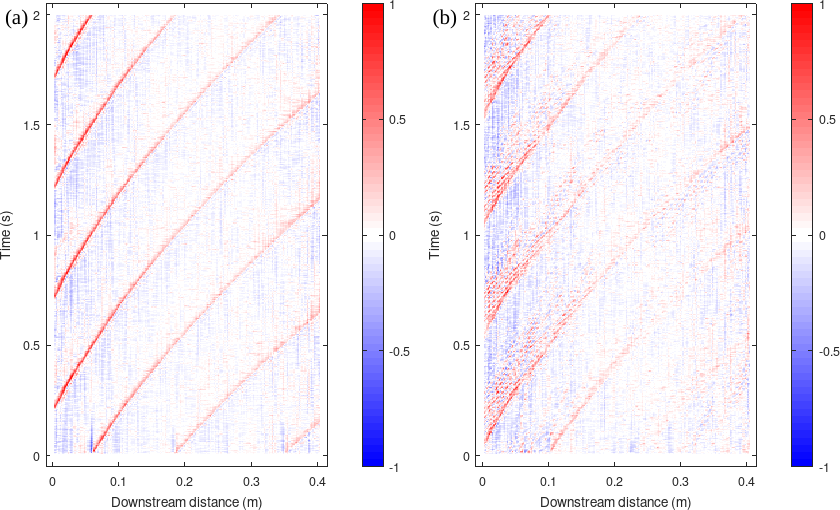}
 \caption{False-colour maps showing the spatiotemporal traces of solitary-like waves forced at the frequency of 2\,Hz. (\textbf{a})~No vibration. (\textbf{b})~The inclined plate is vibrated with the frequency of 48\,Hz and the peak vibration amplitude of 1$g$.
 \label{Fig3}}
\end{figure}

Using our fluorescence intensity analysis software, we register the profiles of the waves at the points located on the centreline of the inclined plate along the downstream direction, and we plot the so-obtained data as a function of time. The resulting spatiotemporal false-colour maps are plotted in Fig.~\ref{Fig3} with the observation period of 2\,s for the scenario of no vibration (Fig.~\ref{Fig3}a) and with the 48\,Hz vibration (Fig.~\ref{Fig3}b). In those figures, we can see the traces of several solitary-like waves that propagate in the downstream direction. The traces are more distinguishable and have a higher false-colour amplitudes in the case of no vibration than with the vibration, which confirms our observation of a decrease in the peak amplitude of the solitary pulses due to the vibration in Fig.~\ref{Fig2}. The ripple waves caused by the vibration-induced Faraday instabilities can also be seen in Fig.~\ref{Fig3}b. It is noteworthy that the separation between the traces and the relative position of the traces in the time-downstream coordinate space are very similar without and with the vibration. This indicates that, even though the peak amplitude of the solitary-like waves is affected by the vibration, in general the vibration does not change the shape of the soliton pulses.

\begin{figure}[H]
 \includegraphics[width=0.7\textwidth]{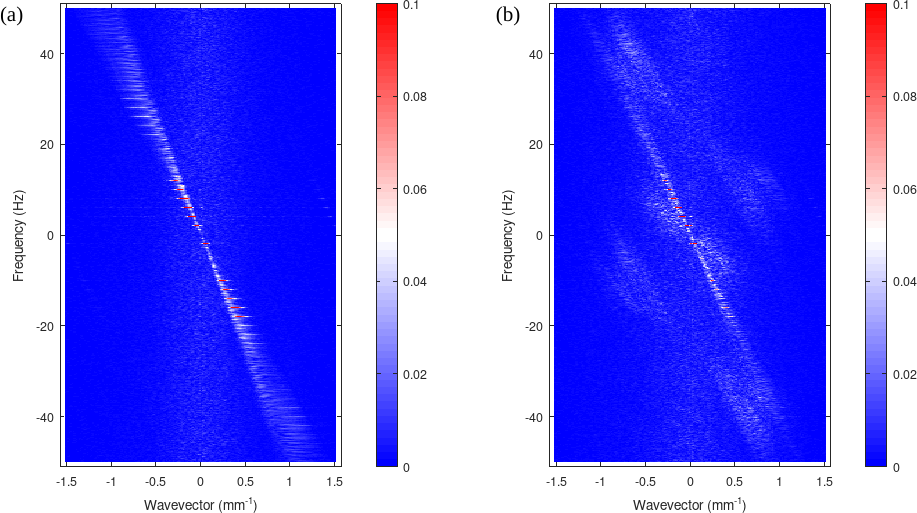}
 \caption{Dispersion maps of the solitary-like waves forced at the frequency of 2\,Hz. (\textbf{a})~No vibration. (\textbf{b})~The inclined plate is vibrated with the frequency of 48\,Hz and the peak vibration amplitude of 1$g$. The plots are slightly oversaturated for the sake of a better visual presentation.
 \label{Fig4}}
\end{figure}

Then, we apply a two-dimensional Fourier transformation to the spatiotemporal data to obtain the dispersion maps as a function of frequency $f$ and wavevector $k$. Since the speed of a wave is given by $u = \omega/k=2\pi f/k$, using the resulting dispersion maps we can identify the bands of constant $f/k$ ratio that correspond to waves travelling along the inclines plate at a constant speed. Yet, we apply the standard f-k filtering procedures to remove noise from the dispersion characteristics \cite{Wan20}. Figure~\ref{Fig4} shows the dispersion maps for the case of no vibration (Fig.~\ref{Fig4}a) and with the 48\,Hz vibration (Fig.~\ref{Fig4}b). While the negative frequency regions of the dispersion maps originate from the mathematical properties of the Fourier transformation, the sign of the wavevector has the physical meaning as it determines the direction of the wave propagation.

\begin{figure}[H]
 \includegraphics[width=0.7\textwidth]{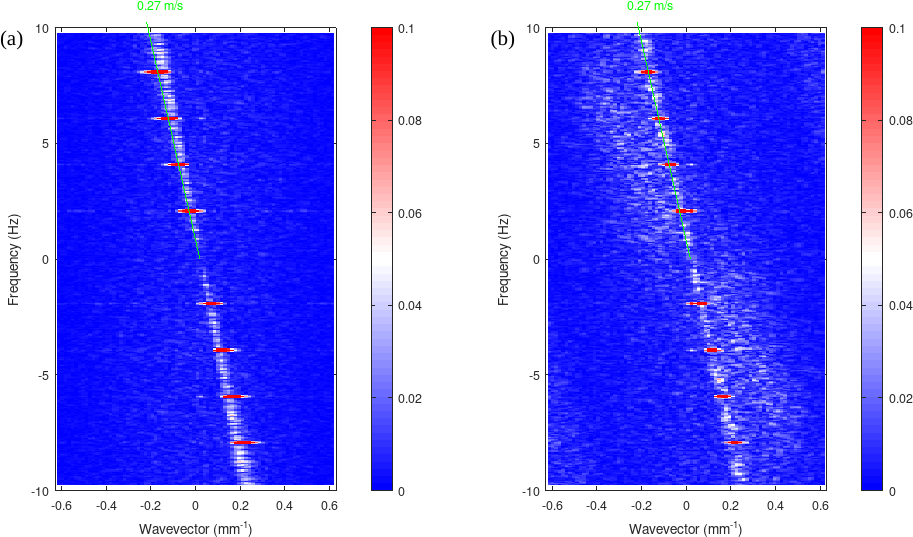}
 \caption{Close-ups of the dispersion maps presented in Fig.~\ref{Fig4}. The linear fits of the dispersion bands and the corresponding wave velocities are shown.
 \label{Fig5}}
\end{figure}

We first analyse the dispersion map in Fig.~\ref{Fig4}a and its zoomed image presented in Fig.~\ref{Fig5}a, where we can see a set of high-magnitude discrete bands that are superimposed on a broader continuum band of a lower magnitude. The frequencies of the discrete bands correspond to the forcing frequency of the solitary-like waves 2\,Hz and its higher-order harmonics of 4, 6 and 8\,Hz and so forth. The origin of the harmonics is due to the nonlinear effects as discussed below. The spectrum of the discrete bands changes as the frequency of forcing of the solitary-like waves is changed. When the modulation of the pump flow was turned off, i.e.~with no wave forcing, the discrete bands completely disappeared. However, a continuum band was always observed independently of whether the forcing was turned on or off. Subsequently, we associate the continuum band with natural periodic rolling waves propagating on the surface of the liquid film flowing over the inclined surface. According to the frequency-wavevector spectral analysis theory \cite{Wan20}, a fit of the observed bands with a straight line produces the velocity of the solitary-like wave of $0.27\pm 0.02$\,m/s.

When the plate is vibrated (Fig.~\ref{Fig4}b), in addition to the dispersion bands discussed in Fig.~\ref{Fig4}a we observe two new isolated bands that can be associated with the Faraday instability. Moreover, the close-up of the dispersion map (Fig.~\ref{Fig5}b) shows that the magnitude of the discrete modes decreased due to the vibration, which is an observation that is consistent with our conclusions made earlier in the text. Yet, the bands in Fig.~\ref{Fig5}b can also be fitted with a straight line that corresponds to the wave velocity of $0.27\pm 0.02$\,m/s.

Empirically, the presence of the discrete bands at the forcing frequency of 2\,Hz and its higher-order harmonics can be explained using the
well-established analogy between the rolling waves and acoustic waves \cite{Ham98}. Indeed, the solitary-like surface waves in Fig.~\ref{Fig2} can be regarded as large-amplitude shock-like disturbances (in the sub-field of physically similar roll waves in open channel such a discontinuity is called the hydraulic jump \cite{Pum83, Bal04, Kal12}). Shock waves are also well-known in the field of nonlinear acoustic, where their formation is accompanied by strong nonlinear effects such as the generation of higher-order harmonic frequencies \cite{Mak19}. Considering longitudinal acoustic waves that can be described as alternating areas of compression and rarefaction in the medium, we can show that the points of the crests of an acoustic wave travel faster than the speed of sound in the medium, but the points of the wave troughs travel slower \cite{Mak19}. This physical process underpins the formation of an acoustic shock wave \cite{Ham98}. In turn, in the field of rolling waves, the crest of a large-amplitude solitary-like wave is connected to its trough by a discontinuity, where the flow regime abruptly changes from a supercritical condition and where the fluid moves faster than the wave, to a subcritical one, where the fluid moves slower \cite{Bal04, Kal12}. As a result, the spectrum of the wave becomes enriched by higher-order harmonic of the frequency of forcing.

Qualitatively similar results were obtained at the vibration frequencies in the range from 30\,Hz to 50\,Hz, and they were validated by our theoretical analysis, the results of which are presented in the following section.

\section{Theory}
The theoretical description of a non-steady flow in the presence of deformable interfaces, such as the flow in a thin liquid layer down a vibrated incline, is a notoriously difficult hydrodynamic problem. The exact analysis is only available in the linear case, i.e.~when the deformation amplitude of the liquid-air interface is much smaller than the average film thickness \cite{Woo95}. In the general case, when a harmonic vibration is applied both in the perpendicular and parallel directions with respect to the the inclined plate, the unperturbed base flow is given by a superposition of a steady Nusselt flow and of an additional harmonically oscillating flow parallel to the incline with a flat free surface \cite{Woo95}. The Navier-Stokes equation for an incompressible fluid can be linearised about the base flow and the Floquet theory-based stability analysis determines if the flow is stable or unstable.

The full nonlinear problem with large amplitude deformation of the film surface can only be studied approximately using simplified hydrodynamic models. Here we use the well-known Shkadov model \cite{Shkadov67,Shkadov68}, which can be derived from the Navier-Stokes equation by assuming a self-similar parabolic longitudinal velocity profile. The model developed by Shkadov has been used earlier to study nonlinear solitary waves in falling liquid films in the absence of vibration \cite{Tri91, Ale85} and to investigate the onset of Faraday waves in vertically vibrated isolated liquid drops \cite{MP19}.

Thus, we consider a liquid film with the local film thickness $h(x,t)$ flowing down an inclined solid plate that makes an angle $\theta$ with the horizontal, as shown in Fig.~\ref{Fig1}. In our model, the $x$-axis is chosen to be parallel to the plate with the positive direction pointing down the incline. To capture rolling waves, we use a one-dimensional version of the Shkadov model, which is formulated as a set of two coupled nonlinear equations for $h(x,t)$ and the local flux across the layer $q(x,t)=\int_0^{h(x,t)}u(x,z,t)\,dz$, where $u(x,z,t)$ is the longitudinal flow velocity and $z$-axis is perpendicular to the incline
\begin{eqnarray}
\label{theq1}
\rho \left[ \partial_t q  +\frac{6}{5}\partial_x \left(\frac{q^2}{h}\right)\right]&=&-\frac{3\mu q}{h^2}+\sigma h\partial_x^3 h-\rho g(t) \cos(\theta)h\partial_x h+\rho g(t)\sin(\theta)h,\nonumber\\
\partial_t h +\partial_x q&=&0,
\end{eqnarray}
where $\mu$ is the dynamic viscosity, $\sigma$ is the liquid-air surface tension and the time-dependent gravity acceleration due to vibration is $g(t)=g(1+a\cos(\omega t))$. The inclination of the plate leads to a re-distribution of the vertical vibration into a longitudinal $g(t)\sin(\theta)$ and an orthogonal $g(t)\cos(\theta)$ components, respectively.

The base flow corresponds to a time-periodic spatially homogeneous flux $q_0(t)$ and a flat film surface $h_0=const$. From Eqs.~(\ref{theq1}) we obtain the following expression by setting $\partial_x q_0(t)=\partial_x h_0=0$:
\begin{eqnarray}
\label{theq2}
q_0(t)=\frac{g\sin(\theta)h_0^3}{3\nu}\left[1+\frac{a\cos(\omega t)}{(2h_0^2/3l_{\text{ac}}^2)^2+1}+\frac{2 h_0^2}{3l_{\text{ac}}^2}\frac{a\sin(\omega t)}{(2h_0^2/3l_{\text{ac}}^2)^2+1}\right],
\end{eqnarray}
where $l_{\text{ac}}=\sqrt{2\nu/\omega}$ represents the length of the acoustic boundary layer.

In the absence of vibration, i.e.~when $a=0$, the base flow is the time-independent Nusselt flow, where the linear stability is well-known in the case of a falling film, i.e.~at $\theta=90^o$ \cite{Tri91, Ale85}. For an arbitrary inclination angle $\theta$, the instability sets in when $\Rey>\cot(\theta)$, where $\Rey=q_0/\nu=\frac{g\sin(\theta)h_0^3}{3\nu^2}$ is the Reynolds number. To put this condition into perspective, for a water film on a $\theta=3^o$ incline, the flow is unstable when $h_0>0.48$\,mm. The corresponding instability is called gravitational instability and it leads to the onset of long surface waves propagating downstream. The wavelength of unstable waves is longer than $\lambda_c=2\pi/k_c$, where $k_c$ is the critical wave vector of the gravitational instability
\begin{eqnarray}
\label{theq3}
k_c=\left( \frac{\rho g\sin(\theta)}{\sigma}\left(\frac{g\sin(\theta)h_0^3}{3\nu^2}-\cot(\theta)\right)\right)^{1/2}.
\end{eqnarray}
Neutrally stable waves with the wavelength $\lambda_c=2\pi/k_c$ propagate downstream with a speed $c$, which is twice as large as the surface speed in the Nusselt flow, i.e.~$c=g\sin(\theta)h_0^2/\nu$.

When the vibration is switched on, the Faraday instability mode develops and it competes with the gravitational instability mode.
The linear stability of a flat film flowing down an incline under the combined action of the longitudinal and orthogonal vibration has been investigated in Ref.~\cite{Bur01} using a theoretical approach based on the exact linearisation of the Navier-Stokes equation \cite{Woo95}. In the relevant work Ref.~\cite{Bes13}, an integral boundary layer model has been formulated and applied to study nonlinear Faraday waves in liquid films on a horizontal plate subjected to horizontal and vertical vibrations. In Refs.~\cite{Gar13, Gar17}, the nonlinear dynamics of a liquid film falling down a vertical vibrated plate is investigated theoretically and experimentally. However, it should be emphasised that large-amplitude surface waves in a liquid film flowing down an incline in the presence of both the longitudinal and orthogonal vibrations have not been studied earlier.

To study the stability of the base flow Eqs.\,(\ref{theq2}) we use the standard plane-wave ansatz $q(x,t)=q_0(t)+\tilde{q}(t)e^{ikx}$ and $h(x,t)=h_0+\tilde{h}(t)e^{ikx}$, where $k$ is the wavevector of the small-amplitude perturbation. By differentiating the second equation in Eqs.\,(\ref{theq1}) with respect to time and the first equation with respect to $x$, the flux perturbation $\tilde{q}$ can be eliminated to yield a complex-valued Mathieu-like equation for the film thickness perturbation $\tilde{h}$
\begin{eqnarray}
\label{theq4}
\partial_{tt} \tilde{h} +A(t)\partial_t \tilde{h}+B(t)\tilde{h}=0,
\end{eqnarray}
with $A(t)=\frac{3\nu}{h_0^2}+\frac{12}{5}ik\frac{q_0(t)}{h_0}$ and $B(t)=ikg(t)\sin(\theta)+\frac{\sigma}{\rho}h_0k^4+g(t)\cos(\theta)h_0k^2+ik6\nu \frac{q_0(t)}{h_0^3}-\frac{6}{5}k^2\frac{q_0(t)^2}{h_0^2}$.

According to the Floquet theory, the solution of Eq.\,(\ref{theq4}) is given by $\tilde{h}(t)=H(t)e^{\lambda t}$, where $H(t)$ is some bounded periodic function with the period $T=2\pi/\omega$ and $\lambda$ is the Floquet exponent. The solution is stable when the real part of the largest Floquet exponent is negative, i.e.~${\text Re}(\lambda)<0$ and it is unstable otherwise. The Floquet exponents are related to the monodromy matrix ${\bm M}$ via ${\text Re}(\lambda)=\frac{1}{T}\ln(|\Lambda|)$, where $\Lambda$ is the eigenvalue of ${\bm M}$. The $2\times 2$ complex-valued monodromy matrix ${\bm M}$ is given by the fundamental solution matrix that is obtained by writing Eq.\,(\ref{theq4}) as a system of two first-order equations and integrating it over one period $T$ with the unit $2\times 2$ matrix as the initial condition.

\begin{figure}[H]
 \includegraphics[width=0.7\textwidth]{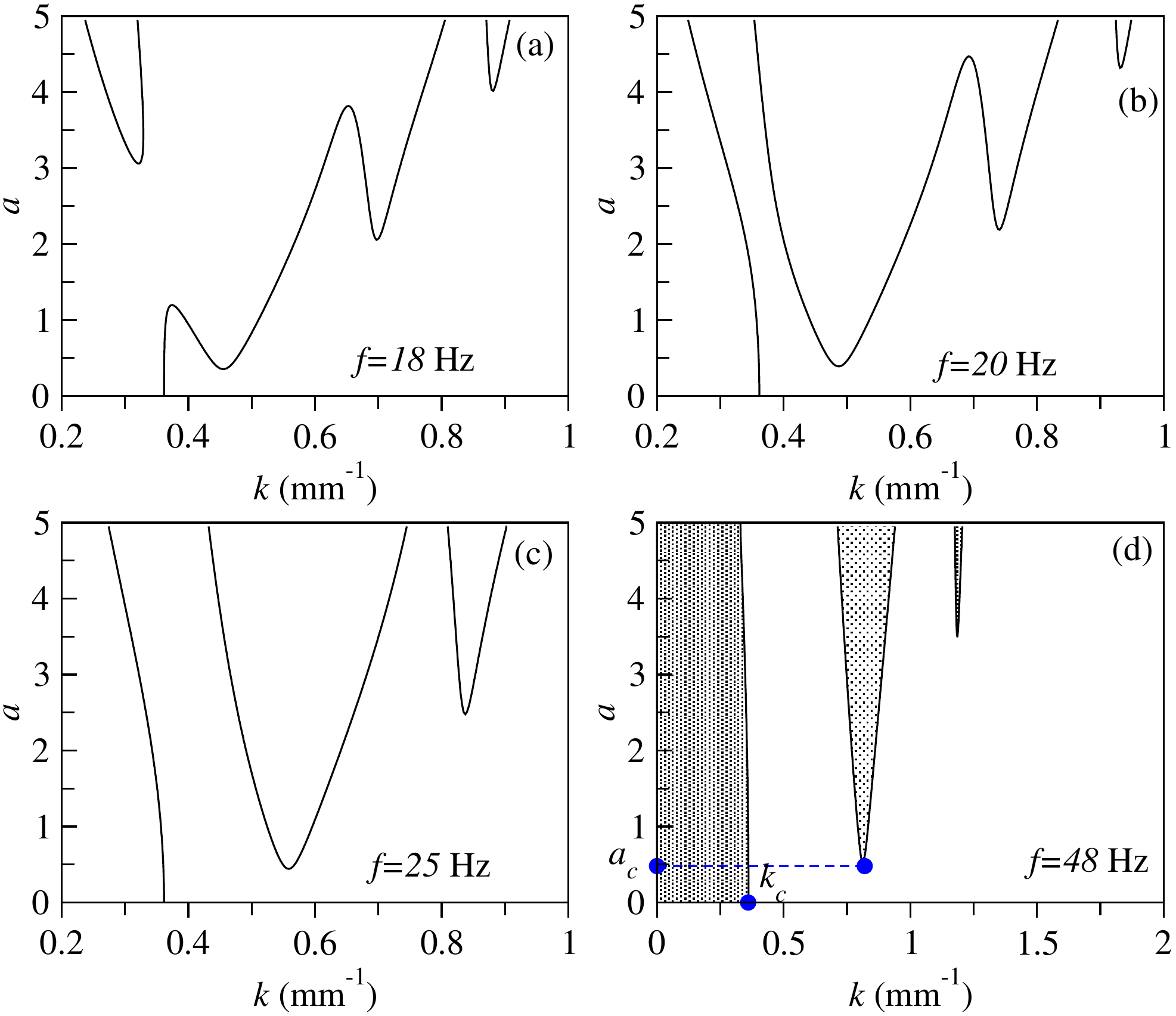}
 \caption{Marginal stability curves of the base flow Eq.\,(\ref{theq2}) for a $0.6$\,mm water film on a $\theta=3^0$ incline vibrated at (\textbf{a})~$f=18$\,Hz, (\textbf{b})~$f=20$\,Hz, (\textbf{c})~$f=25$\,Hz and (\textbf{d})~$f=48$\,Hz. The shaded regions in panel~(\textbf{d}) highlights the unstable areas, $k_c$ indicates the critical wave vector of the gravitational long-wave instability Eq.\,(\ref{theq3}) and $a_c$ marks the critical vibration amplitude when the Faraday instability sets in.
 \label{Fig_th1}}
\end{figure}

For the inclination angle $\theta=3^0$, we choose the thickness of the water film $h_0=0.6$ mm, which is slightly above the critical value for the gravitational instability of $h_0=0.48$\,mm. The marginal stability curves that correspond to ${\text Re}(\lambda)=0$ are shown in Fig.~\ref{Fig_th1} for four different vibration frequencies $f=18,20,25,48$ Hz. The critical wavevector of the gravitational instability Eq.\,(\ref{theq3}) is marked by $k_c$ in Fig.~\ref{Fig_th1}d and it remains unaffected by the vibration. The shaded regions in Fig.~\ref{Fig_th1}d indicate the unstable areas. The Faraday instability sets in at the vibration amplitude $a_c$ that corresponds to the tip of the lowest Faraday tongue. The value of $a_c$, as extracted from Fig.~\ref{Fig_th1}, slightly increases with $f$, namely: $a_c=0.33$ for $f=18$\,Hz, $a_c=0.35$ for $f=20$\,Hz, $a_c=0.38$ for $f=25$\,Hz and $a_c=0.48$ for $f=48$\,Hz. This observation confirms the earlier statement that, for the range of frequencies between 30\,Hz and 50\,Hz, the surface waves are much more sensitive to the changes of the vibration amplitude $a$ than to the changes of the vibration frequency $f$. Indeed, comparing Figs.~\ref{Fig_th1}c,d we see only a marginal difference in the critical amplitude $a_c$ when the frequency is doubled. On the other hand, increasing the value of $a$ from $a=0.5$ to $a=1$ will significantly broaden the band of unstable wavevectors of the Faraday instability, thus significantly changing the dynamics of the surface waves.

To better understand the temporal signature of the surface waves in response to vibration, we compute the imaginary part of the Floquet exponent ${\text Im}(\lambda)=\frac{\omega}{2\pi}({\text arg}(\Lambda))+\omega n$, where, as before, $\Lambda$ is the eigenvalue of the monodromy matrix and $n$ is an arbitrary integer. Any neutrally stable wave, i.e.~${\text Re}(\lambda)=0$, can be represented in the form $\tilde{h}(x,t)=e^{ikx}H(t)e^{\lambda t}=H(t)e^{ikx+i{\text Im}(\lambda)t}$, where $H(t)$ is a bounded $2\pi/\omega$-periodic function.
Therefore, the temporal spectrum of such a neutrally stable wave contains delta-peaks located at $\frac{\omega}{2\pi}({\text arg}(\Lambda))+\omega n$. The temporal spectrum of a growing wave with ${\text Re}(\lambda)>0$ contains the same delta peaks that will appear slightly smeared.

At this stage, it is important to emphasise that the temporal response of the surface waves that develop on the surface of a liquid layer on a vibrated incline is not necessarily harmonic (frequencies $\omega n$) or subharmonic (frequencies $\omega/2+\omega n$). This feature is in stark contrast to the standard Faraday instability in horizontal liquid layers, when the neutrally stable waves are always harmonic or subharmonic standing waves \cite{Kum96}. In some special cases, however, such as discussed in Ref.~\cite{Gar17} for transversally vibrated falling liquid films, the magnitude of $\frac{\omega}{2\pi}({\text arg}(\Lambda))$ may be close to zero or $\omega/2$, leading to an almost harmonic or subharmonic response. For the fluid parameters used in the present study, the frequency of the Faraday mode is significantly shifted from $\omega$ or $\omega/2$, as shown in Fig.~\ref{Fig4}b.

\begin{figure}[H]
 \includegraphics[width=0.6\textwidth]{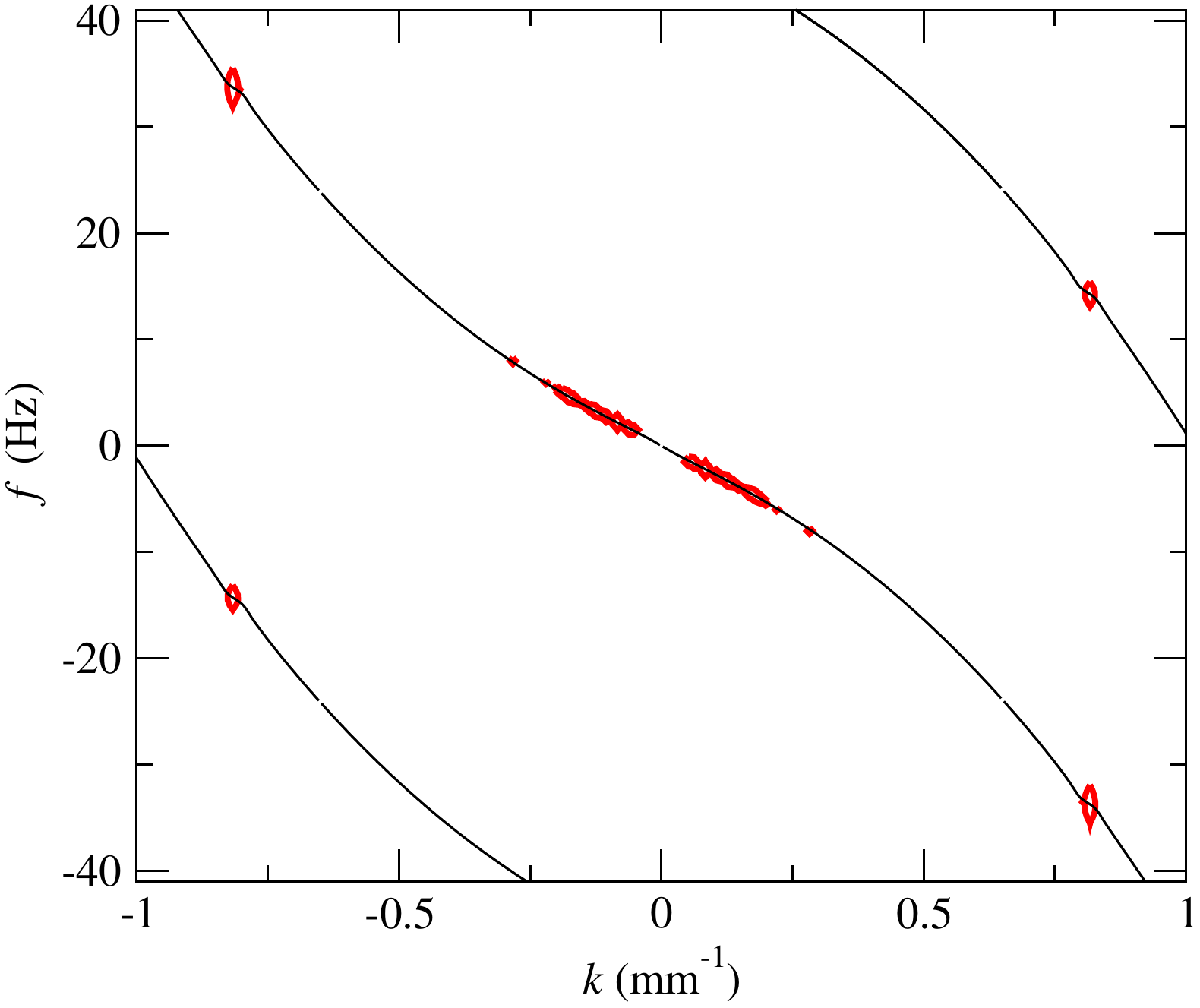}
 \caption{Contour plot (thick lines) of the dispersion map of the solution of Eqs.\,(\ref{theq1}) computed over the time interval of $3$\,seconds with a random initial perturbation of the flat film surface. The thickness of the water film is $h=0.6$\,mm and the vibration parameters are $a=0.8$ and $f=48$\,Hz, similarly to Fig.~\ref{Fig4}b. The thin solid lines correspond to the imaginary part of the Floquet exponent ${\text Im}(\lambda)$ of the most unstable mode.
 \label{Fig_th2}}
\end{figure}

Next, we simulate the experimental conditions, at which the results shown in Fig.~\ref{Fig4}b was obtained, to gain a better understanding of how the vibration changes the dynamics of the waves in the early stages of evolution. Thus, we numerically integrate Eqs.\,(\ref{theq1}) over the time interval of $3$\,seconds in the system of length of $60$\,cm with periodic boundaries. The vibration amplitude is $a=0.8$ and the other parameters are the same as in Fig.~\ref{Fig_th1}d. As the initial conditions, we use zero flux and random initial perturbation of the flat film surface with the amplitude of $10^{-3}$\,mm. The dispersion map is obtained taking the two-dimensional Fourier transformation of the solution $h(x,t)$. The contour lines of the dispersion map that correspond to the level of 3\% of its maximum are shown by the thick lines in Fig.\,\ref{Fig_th2}. The thin solid lines in Fig.\,\ref{Fig_th2} correspond to the dispersion curves ${\text Im}(\lambda)(k)$, computed from Eq.\,(\ref{theq4}) for $a=0.8$ and $f=48$\,Hz. It can be seen that the results of the direct simulation of the full system Eq.\,(\ref{theq1}) are in perfect agreement with the dispersion curves of the small-amplitude surface waves.

The dispersion map in Fig.\,\ref{Fig_th2} is dominated by the delta-peaks located at $f=\pm 14$ and $f=\pm34$\,Hz, thus confirming that the primary response of the liquid film to a harmonic vibration is neither harmonic, nor subharmonic. Qualitatively, the shift of the response frequency away from the standard for the Faraday instability subharmonic mode can be explained as follows. In a horizontal layer vibrated at frequency $\omega$, the Faraday instability sets in the form of a standing wave oscillating at the subharmonic frequency $\omega/2$. Any standing wave can be represented as a superposition of two plane waves travelling at the phase speed of $c=\omega/(2k)$ in the opposite directions, i.e.~$h(x,t)=Ae^{i\omega t/2+ikx}+Ae^{i\omega t/2-ikx}+cc$. When the layer is slightly inclined with the positive direction pointing downstream, it would be reasonable to assume that the plane wave propagating downstream will increase its phase speed by some amount $\delta c$, but the wave propagating upstream will decrease its phase speed by the same amount $\delta c$. Assuming that the wavevector remains unaffected by a small inclination angle, the resulting solution is represented by $h(x,t)=Ae^{i(\omega/2-k\delta c) t/2+ikx}+Ae^{i(\omega/2 +k\delta c )t/2-ikx}+cc$. Therefore, the temporal spectrum of $h(x,t)$ will contain delta-peaks located at $\pm(\omega/2 +k\delta c)$ and $(\pm (\omega/2-k\delta c))$, in agreement with Fig.\,\ref{Fig_th2}.

Alongside the delta-peaks, the dispersion map in Fig.\,\ref{Fig_th2} also contains a band of linearly unstable plane waves with the wavevectors $k<k_c$. These long waves are amplified as the result of the gravitational instability mode. It can be seen that the gravitational band falls perfectly on the central dispersion curve that passes through the origin. The central dispersion branch in Fig.\,\ref{Fig_th2} is almost indistinguishable from the dispersion curve in the absence of vibration (not shown). This allows us to conclude that a relatively strong vibration (sufficiently strong to excite Faraday waves) has almost no effect on the phase speed $c={\text Im}(\lambda)/k$ of the long gravitational surface waves.

\begin{figure}[H]
 \includegraphics[width=0.7\textwidth]{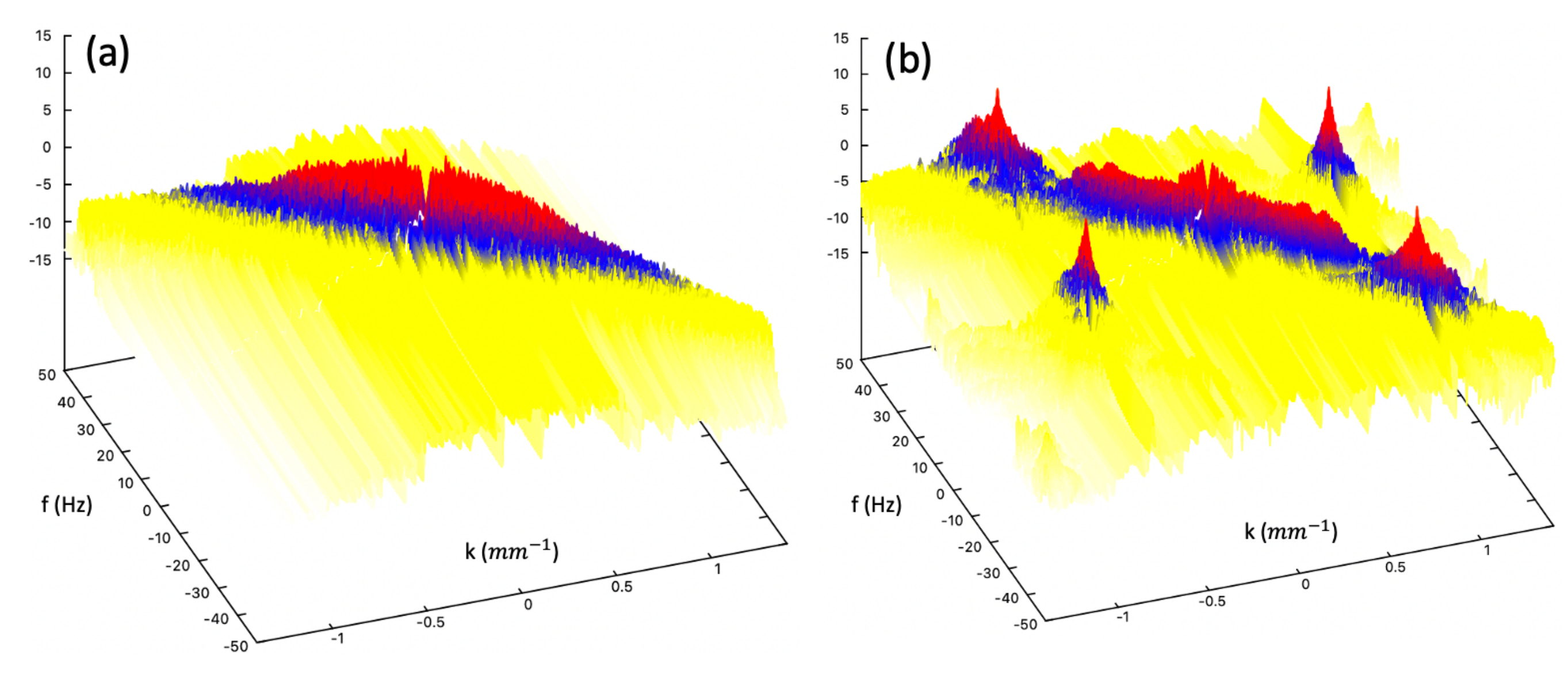}
 \caption{(\textbf{a})~Dispersion map obtained from the solution of Eqs.\,(\ref{theq1}) in the absence of vibration for a $0.6$\,mm water film on a $\theta=3^o$ incline. (\textbf{b})~Dispersion map of the solution of Eqs.\,(\ref{theq1}) when the inclined plane vibrated at $f=48$\,Hz with the amplitude $a=0.8$. The scaling for the vertical axis is in arbitrary logarithmic units.
 \label{Fig_th3}}
\end{figure}

To study the interaction between the Faraday waves and gravitational surface waves in the nonlinear regime, we solve Eqs.\,(\ref{theq1}) over the time interval of $15$\,seconds with and without vibration and compare the respective dispersion maps in Fig.~\ref{Fig_th3}.

It is evident from Fig.~\ref{Fig_th3} that vibration leads to a suppression of the long surface waves. Indeed, the magnitude of the dispersion band that corresponds to the gravitational waves is significantly smaller when the film is vibrated. This result is in qualitative agreement with Fig.~\ref{Fig2}.

\section{Conclusions}
In conclusion, our experiments with a sub-millimetre thick water layer on a slightly inclined vertically vibrated plate demonstrate that low-frequency vibration in the range between 30 and 50\,Hz suppresses the development of long rolling surface waves propagating downstream. These surface waves appear as the result of the long-scale gravitational instability of the base flow in the absence of vibration \cite{Yih63, Ben66} and may also be excited by mechanically perturbing the flow at the inlet. A relatively small thickness of the water layer (under 1\,mm) is required to suppress the three-dimensional instability of the rolling waves that is known to develop at large flow rates. Experimental findings are verified using a boundary-layer hydrodynamic model \cite{Shkadov67,Shkadov68} obtained from the Navier-Stokes equation by assuming a self-similar parabolic longitudinal flow velocity. Linear stability and nonlinear dynamics of the surface waves obtained with the model qualitatively confirm the main experimental findings.

Without vibration, the Fourier content of surface waves is represented by a broad band of unstable wave vectors $k<k_c$, where $k_c$ is a critical cut-off wave vector of the gravitational instability (see Eq.~\ref{theq3}). As the instability unfolds, the wavelength of the dominant wave quickly increases until it develops into a solitary-like wave \cite{Gol94}. For fluids with a relatively small viscosity, such as water, the characteristic time required for solitary rolling waves to develop on a $3^o$ incline is of the order of several seconds. In the nonlinear regime, the Fourier content of the surface waves is dominated by solitary-like waves characterised by a small wavevector with a background of smaller amplitude shorter waves, which is shown in Fig.~\ref{Fig_th3}a and Fig.~\ref{Fig4}a.

We observe that the properties of the surface waves change dramatically when the layer is vibrated. Thus, a relatively weak vibration (the vibration amplitude $a<g$) leads to the onset of the secondary Faraday instability in the form of short waves with a wavelength of $\lambda\approx 5\dots 10$\,mm when vibrated at $f=48$\,Hz. In agreement with the earlier theoretical studies \cite{Woo95, Bur01, Gar13}, the temporal frequency of the Faraday waves is shifted away from the harmonic ($48$\,Hz) and subharmonic ($24$\,Hz) response that is typical of Faraday instability in horizontal liquid layers. In fact, the inclination angle of the plate acts as a wave filter, splitting a standing Faraday wave into two plane waves: one propagating upstream and one propagating downstream. Similarly to the Doppler effect, the wave that propagates downstream increases its speed and, therefore, increases its temporal frequency, while the wave that propagates upstream decreases its speed and frequency. For water layers vibrated at $48$\,Hz, we observe the following shifts in frequency away from the subharmonic response: from $24$\,Hz to approximately $40$\,Hz for the downstream wave and from $24$\,Hz to approximately $8$\,Hz for the upstream wave.

In the nonlinear regime, the interaction between shorter Faraday waves and longer gravitational waves leads to the broadening of their respective bands in the f-k dispersion map. Most importantly, we find that the average and peak amplitudes of the long-scale gravitational waves are significantly reduced when vibration is applied. This result is rather intriguing since the total influx of energy is larger in the vibrated system when both gravity and vibration together drive the flow, unlike in the non-vibrated case, where the only source of energy is due to gravity. Yet, nonlinear wave interaction leads to an uneven re-distribution of energy amongst the Faraday and gravitational waves in favour of the former. The physical mechanism responsible for the suppression of gravitational waves remains an open question; however, it is plausible to assume that the fast-oscillating fluid flow in the form of circulation patterns \cite{Perinet2017} in pulsating Faraday waves may slow down the redistribution of fluid on the large scale, required for the growth and development of the gravitational waves. This result is even more surprising since we did not observe any noticeable change in the velocity of the gravitational waves induced by vibration.

Apart from a contribution of the fundamental knowledge, the results presented in this work may be used to better understand and further improve certain technological processes that rely on falling liquid films. Yet, the demonstrated immunity of the solitary-like waves to external vibration and their intriguing nonlinear dynamical behaviour will be of interest to researchers working on emergent technologies, where both solitary waves and fluidic systems play an important role \cite{Hei05, Gon14, Che17, Tan19, Sil21, Mat22, Mak22}.

%%%%%%%%%%%%%%%%%%%%%%%%%%%%%%%%%%%%%%%%%%
\vspace{6pt}

%%%%%%%%%%%%%%%%%%%%%%%%%%%%%%%%%%%%%%%%%%
%% optional
%\supplementary{The following supporting information can be downloaded at:  \linksupplementary{s1}, Figure S1: title; Table S1: title; Video S1: title.}


%%%%%%%%%%%%%%%%%%%%%%%%%%%%%%%%%%%%%%%%%%
\authorcontributions{I.S.M. conducted the experiments. A.P. conducted the theoretical analysis. Both authors wrote the manuscript.}

%\funding{Not applicable.}

%\institutionalreview{Not applicable.}

%\informedconsent{Not applicable.}

%\dataavailability{We encourage all authors of articles published in MDPI journals to share their research data. In this section, please provide details regarding where data supporting reported results can be found, including links to publicly archived datasets analyzed or generated during the study. Where no new data were created, or where data is unavailable due to privacy or ethical re-strictions, a statement is still required. Suggested Data Availability Statements are available in section “MDPI Research Data Policies” at \url{https://www.mdpi.com/ethics}.}

%\acknowledgments{In this section you can acknowledge any support given which is not covered by the author contribution or funding sections. This may include administrative and technical support, or donations in kind (e.g., materials used for experiments).}

\conflictsofinterest{The authors declare no conflict of interest.}

\end{paracol}
%%%%%%%%%%%%%%%%%%%%%%%%%%%%%%%%%%%%%%%%%%
%\begin{adjustwidth}{-\extralength}{0cm}

\reftitle{References}


\externalbibliography{yes}
\bibliography{refs}

% If authors have biography, please use the format below
%\section*{Short Biography of Authors}
%\bio
%{\raisebox{-0.35cm}{\includegraphics[width=3.5cm,height=5.3cm,clip,keepaspectratio]{Definitions/author1.pdf}}}
%{\textbf{Firstname Lastname} Biography of first author}
%
%\bio
%{\raisebox{-0.35cm}{\includegraphics[width=3.5cm,height=5.3cm,clip,keepaspectratio]{Definitions/author2.jpg}}}
%{\textbf{Firstname Lastname} Biography of second author}

% For the MDPI journals use author-date citation, please follow the formatting guidelines on http://www.mdpi.com/authors/references
% To cite two works by the same author: \citeauthor{ref-journal-1a} (\citeyear{ref-journal-1a}, \citeyear{ref-journal-1b}). This produces: Whittaker (1967, 1975)
% To cite two works by the same author with specific pages: \citeauthor{ref-journal-3a} (\citeyear{ref-journal-3a}, p. 328; \citeyear{ref-journal-3b}, p.475). This produces: Wong (1999, p. 328; 2000, p. 475)

%%%%%%%%%%%%%%%%%%%%%%%%%%%%%%%%%%%%%%%%%%
%% for journal Sci
%\reviewreports{\\
%Reviewer 1 comments and authors’ response\\
%Reviewer 2 comments and authors’ response\\
%Reviewer 3 comments and authors’ response
%}
%%%%%%%%%%%%%%%%%%%%%%%%%%%%%%%%%%%%%%%%%%
%\PublishersNote{}
%\end{adjustwidth}
\end{document}